\documentclass[prl,twocolumn,preprintnumbers,superscriptaddress,amsmath,amssymb]{revtex4}

\usepackage{graphicx}
\usepackage{dcolumn}
\usepackage{bm}
\usepackage{natbib}
\usepackage{amsmath}
\usepackage{amssymb}
\usepackage{float}
\usepackage{datetime}

\usepackage[normalem]{ulem}

\begin{document}

\author{Stefano Roddaro}
\email{s.roddaro@sns.it}
\affiliation{NEST, Scuola Normale Superiore and Istituto Nanoscienze-CNR, Piazza S. Silvestro 12, I-56127 Pisa, Italy}
\affiliation{Istituto Officina dei Materiali -- CNR, Basovizza S.S. 14 km 163.5, I-34149 Trieste, Italy}
\author{Daniele Ercolani}
\affiliation{NEST, Scuola Normale Superiore and Istituto Nanoscienze-CNR, Piazza S. Silvestro 12, I-56127 Pisa, Italy}
\author{Mian Akif Safeen}
\affiliation{NEST, Scuola Normale Superiore and Istituto Nanoscienze-CNR, Piazza S. Silvestro 12, I-56127 Pisa, Italy}
\author{Soile Suomalainen}
\affiliation{Optoelectronics Research Centre, Tampere University of Technology, P.O. Box 692,
FIN-33101 Tampere, Finland}
\author{Francesco Rossella}
\affiliation{NEST, Scuola Normale Superiore and Istituto Nanoscienze-CNR, Piazza S. Silvestro 12, I-56127 Pisa, Italy}
\author{Francesco Giazotto}
\affiliation{NEST, Scuola Normale Superiore and Istituto Nanoscienze-CNR, Piazza S. Silvestro 12, I-56127 Pisa, Italy}
\author{Lucia Sorba}
\affiliation{NEST, Scuola Normale Superiore and Istituto Nanoscienze-CNR, Piazza S. Silvestro 12, I-56127 Pisa, Italy}
\author{Fabio Beltram}
\affiliation{NEST, Scuola Normale Superiore and Istituto Nanoscienze-CNR, Piazza S. Silvestro 12, I-56127 Pisa, Italy}

\title{Giant thermovoltage in single InAs-nanowire field-effect transistors}

\begin{abstract}

Millivolt range thermovoltage is demonstrated in single InAs-nanowire based field effect transistors. Thanks to a buried heating scheme, we drive both a large thermal bias $\Delta T>10\,{\rm K}$ and a strong field-effect modulation of electric conductance on the nanostructures. This allows the precise mapping of the evolution of the Seebeck coefficient $S$ as a function of the gate-controlled conductivity $\sigma$ between room temperature and $100\,{\rm K}$. Based on these experimental data a novel estimate of the electron mobility is given. This value is compared with the result of standard field-effect based mobility estimates and discussed in relation to the effect of charge traps in the devices.

\end{abstract}



\maketitle

The quest for large-scale solid-state thermoelectric (TE) conversion was one of the major driving forces behind semiconductor research before the discovery of the transistor effect~\cite{Majumdar04}. Expectations were not met, however, largely because of efficiency issues and the high costs and material toxicity. Indeed achievement of efficient TE devices requires a non-trivial tuning of interdependent material parameters and can be expressed in terms of the maximization of the figure of merit $ZT=\sigma S^2T/\kappa$, where $S$ is the Seebeck coefficient, $\sigma$ and $\kappa$ are the electrical and thermal conductivities, $T$ the average operation temperature~\cite{Majumdar04,BookTE}. Large $ZT$ values proved elusive over the past decades despite the great design flexibility offered by semiconductor heterostuctured materials~\cite{Snyder08}. Nanotechnology has recently revived the interest on this research area and may be able to play the role of a game-changer. It enables the design of artificial materials with novel properties and thus opens alternative routes to the optimization of thermodynamic efficiency~\cite{Dresselhaus07,Vineis10,Shi12,Venkatasubramanian01,Boukai08,Poudel08,Heremans08}. Among these, nanowire-based structures have attracted a significant interest~\cite{Shi10,Martinez11,Lee11,Zhou11,Tian12,Moon13}.

\begin{figure}[h!]
\begin{center}
\includegraphics[width=0.5\textwidth]{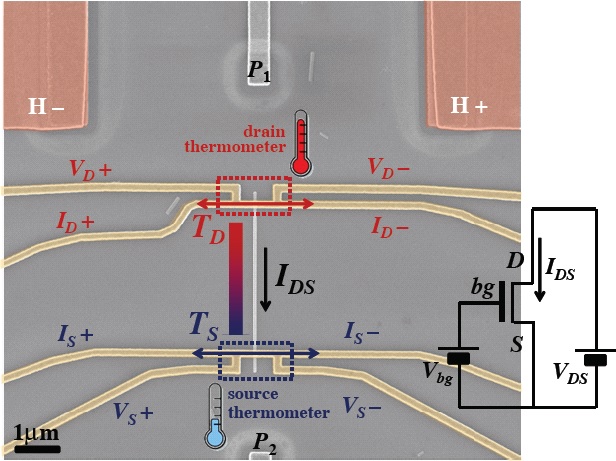}
\caption{Scanning electron micrograph of one of the studied devices: an InAs nanowire on a ${\rm SiO_2/Si}$ substrate is connected to two resistive Ti/Au thermometers (yellow). A strong thermal gradient is induced using a buried heater (contacts visible in pink). The set-up allows the simultaneous (i) determination of the $IV$ characteristics of the nanowire, the (ii) induction and (iii) measurement of different temperatures $T_1$ and $T_2$ at the two contacts. The set-up allows the simultaneous determination of the Seebeck coefficient $S$ and resistance $R$ of the nanowire and the tuning of these parameters by field effect.}
\end{center}
\end{figure}

In this Letter, we focus on the TE properties of single InAs nanowires (NWs), self-assembled nanostructures actively investigated in view of a number of potential applications in novel electronics~\cite{RoddaroNR11,Spathis11,Giazotto11} and optoelectronics~\cite{Vitiello12,Pitanti11}. The present InAs NW model system is also particularly relevant for the investigation of strongly-confined low-dimensional systems~\cite{ThelanderAPL2003, Bjork04, Fuhrer07, Roddaro11, Romeo12}, which in turn may make available exciting heat-transport and TE properties~\cite{Hoffmann09,Zhang11}. Here we demonstrate that record-high thermovoltage values in excess of $1\,{\rm mV}$ can be induced in devices comprising a single-NW as active element {\em and} that these values can be modulated by field-effect. A buried-heater approach allows us to combine a large thermal gradient with the field-effect control of single nanostructures deposited on standard oxidized silicon. We highlight that this scheme can be of general interest for the investigation of a large class of nanomaterials with potential TE applications. We shall provide a precise map of the Seebeck coefficient in our NWs as a function of $\sigma$ from room temperature down to $\approx100\,{\rm K}$. Also, since $S$ vs. $\sigma$ curves are strongly dependent on electron scattering times, our experimental data lead to an estimate of the electron mobility $\mu_{e,S}$, which we shall compare with the values obtained with the standard approach based on field-effect mobility $\mu_{e,F\!E}$. The role of surface states and gate hysteresis~\cite{Halpern12} in NW transport properties and in our parameter estimates will be discussed. 

One of the single-NW field-effect transistors (FETs) studied is depicted in Fig.~1. Devices were built starting from $73\pm7\,{\rm nm}$-diameter InAs NWs deposited by drop casting over the SiO$_2$/Si substrate. Source (S) and drain (D) electrodes were fabricated at different distances ranging from $1\,{\rm \mu m}$ up to $3.5\,{\rm \mu m}$ (device in the figure). These contacts could be operated both as electrical leads and resistive thermometers, thanks to their multi-contact arrangement. Consequently we could sample temperatures $T_S$ and $T_D$ at the two ends of the FET and the thermal bias $\Delta T=T_D-T_S$. The four-wire thermometers were operated using a small AC current bias ($\approx 5\,{\rm \mu A}$) and a phase-locked measurement technique while the DC transport along the NW was being simultaneously monitored. The doped Si substrate played the dual role of backgate electrode and differential heater. Two current injection leads $H+$ and $H-$ were used to establish a non-uniform current density in the substrate portion below the NW. By suitably tuning heater bias $V_{H\pm}$ we could control both the temperature jump along the wire ($T_D$ and $T_S$ were measured using the resistive thermometers) and the local backgate potential $V_{bg}$. Operation details and performance of the present heating scheme will be presented elsewhere, while further information on the devices and the measurement procedures are provided in the Methods and in the Supplementary Material sections.

As shown in Fig.~2a, the IV characteristics of our FETs were always linear within the explored electrical/thermal biasing ranges according to:

\begin{equation}
R	\cdot I_{DS} = V_{DS}+V_{th}
\end{equation}

\begin{figure}[h!]
\begin{center}
\includegraphics[width=0.50\textwidth]{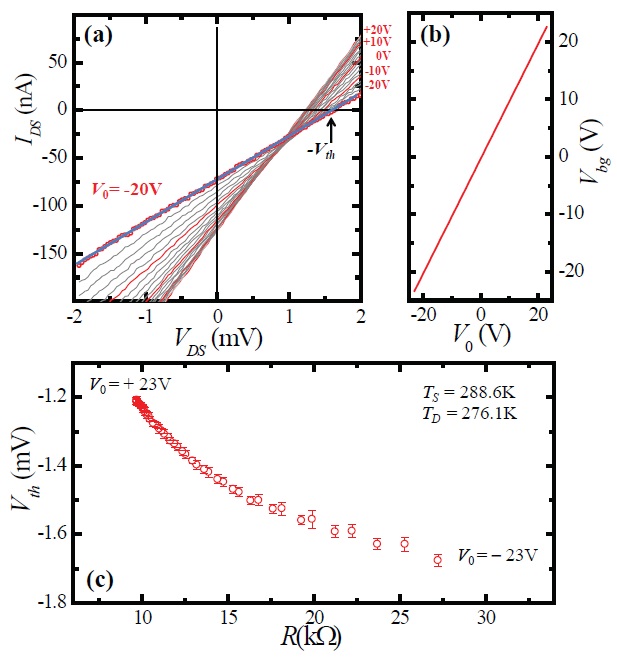}
\caption{(a) IV curves as a function of the $V_0$ parameters for a thermal bias of $\Delta T\approx12.5\,{\rm K}$. Curves are carefully calibrated in order to eliminate measurement offsets, so that no current or voltage shift is observed in isothermal conditions $\Delta T=0$. The intercept at $V_{DS}=0$ allows to extract the thermovoltage $V_{th}$. (b) Measured backgate voltage as a function of $V_0$: a small deviation ($-0.377\,{\rm V}$ in the plot) is typically observed due to the non-ideal behavior of the $H_\pm$ contacts. (c) A large thermovoltage $V_{th}$ can be obtained and controlled by field effect. This evolution can be directly linked to a change of resistance $R$ in the nanowire.}
\end{center}
\end{figure}

\noindent where $R$ is the NW resistance and $V_{t\!h}=(S-S_{\rm Au})\Delta T$ is the thermovoltage due to the NW-Au thermocouple, $S$ and $S_{\rm Au}$ are the Seebeck coefficients of the NW and of the electrodes, respectively. Very large thermovoltages beyond $-1\,{\rm mV}$ were obtained and a strong modulation could be driven by field effect. The FET behavior was measured at temperatures ranging from $100\,{\rm K}$ up to $300\,{\rm K}$ (values were calculated as the average between $T_S$ and $T_D$). Differential heating was obtained by applying a heater bias $V_{H\pm}=V_0\pm V_H$ with respect to the device ground contact $S$. This led at the same time to a backgate potential $V_{bg}\approx V_0$. The gating bias was also directly monitored by probes $P1$ and $P2$ in order to correct for any deviations from device symmetry. For every heating configuration (i.e. for a given $V_H$ value) a full backgate scan was performed stepping $V_0$ from $-23\,{\rm V}$ to $+23\,{\rm V}$ with a relatively slow speed of about $1\,{\rm V/min}$. At each gating value, an IV sweep was recorded in alternating directions, along with the back-gate voltage value $V_{bg}(V_0)$. Figure~2a shows the IV curves measured for the device in Fig. 1 with $V_H=1.8\,{\rm V}$ which yielded $T_S=288.6\,{\rm K}$ and $T_D=276.1\,{\rm K}$. The correspondence between $V_0$ and the backgate voltage $V_{bg}$ is shown in Fig.~2b: for this $V_H$ value $V_{bg}\approx V_0-0.377\,{\rm V}$, the small difference being linked to the resistance drops at the $H_\pm$ contacts. From these data we can readily calculate thermovoltage $V_{t\!h}(V_0)$ and resistive slope $R(V_0)$. Figure~2c shows the resulting evolution of thermovoltage $V_{th}$ versus resistance $R$.

Thanks to the direct measurement of temperatures $T_D$ and $T_S$ and to the large $\Delta T$ attainable, a rather precise mapping of $S(\sigma,T)$ was determined for over 20 temperature values between $100\,{\rm K}$ and $320\,{\rm K}$ and for $\Delta T$ ranging from a minimum of $\approx0.5\,{\rm K}$ to a maximum of $\approx12.5\,{\rm K}$. Figure~3a reports some of the curves obtained. Note that the small contribution due to $S_{\rm Au}$ was subtracted using $S_{\rm Au}=T\cdot 6.4\,{\rm nV/K^2}$, a good approximation in the explored temperature regime~\cite{Mott}. In addition, $\sigma$ was calculated as $L/((R-R_s)\cdot\Sigma)$, where $R_s=1\,{\rm k\Omega}$ is the known series resistance of the measurement set-up (contact resistances are negligible with respect to $R$) and $L=2.6\pm 0.1\,{\rm \mu m}$. The hexagonal NW section $\Sigma$ was calculated as $\Sigma=3\sqrt{3}\cdot r^2/2=1.38\pm0.27\times10^4\,{\rm nm^2}$, given $2r=73\pm 7{\rm nm}$ is the corner-to-corner ``diameter'' that was measured by scanning electron microscopy. The corresponding power factor $S^2\sigma$ is plotted in Fig.3b and peaks on the high-conductivity side of the graph. We also note that $S$ displays a monotonic dependence on $T$ for every given value of $\sigma$ with an almost linear dependence on $T$. This is directly visible in Fig.~3c where $S/T$ values show a good overlap and the slight sublinearity of $S$ vs $T$ becomes evident.

The overall evolution of the NW Seebeck coeffiecient can be understood starting from the approximate law

\begin{equation}
S \approx -\frac{\pi^2k_B^2T}{3e}\cdot \frac{1}{\sigma}\cdot\left.\frac{d\sigma}{dE}\right|_\mu
\end{equation}

\begin{figure}[h!]
\begin{center}
\includegraphics[width=0.4\textwidth]{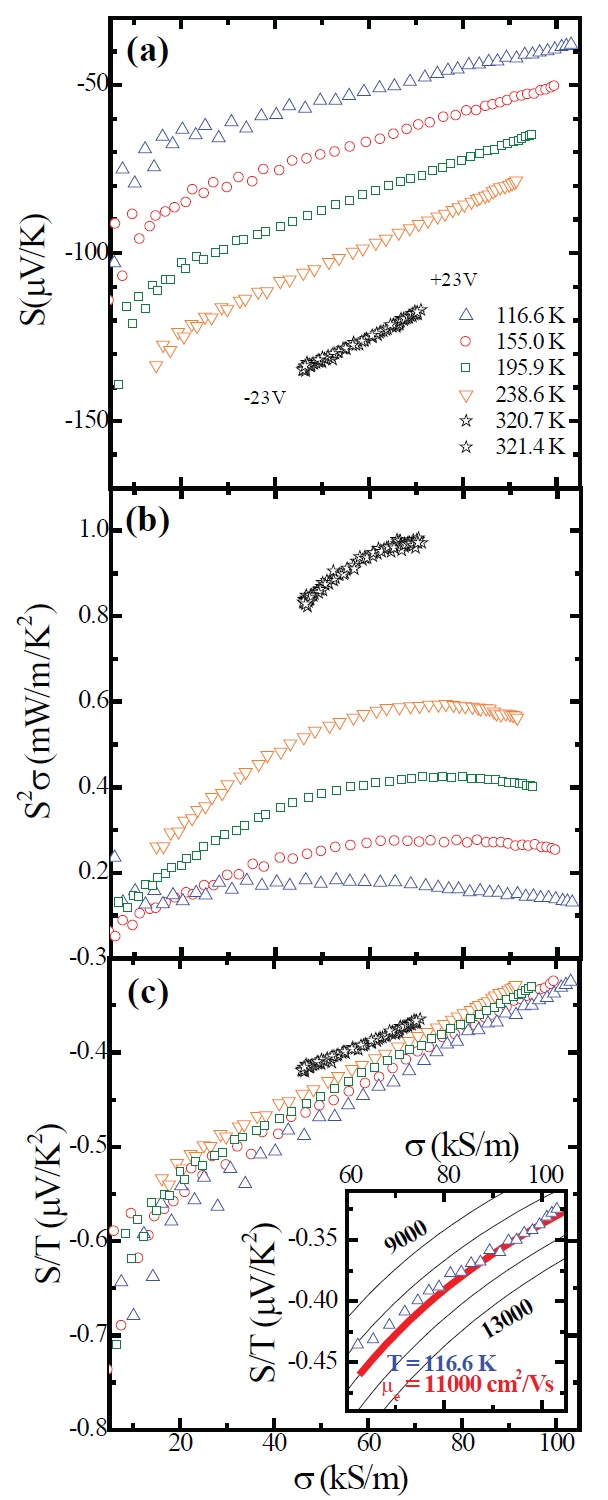}
\caption{Panel (a): selected curves $S(\sigma)$ for one of the studied devices at temperatures $T=116.6$, $155.0$, $195.9$, $238.6$, $320.7$ and $321.4\,{\rm K}$. The corresponding power factor $S^2\sigma$ is shown in panel (b). The $T$ dependence of $S$ is slightly sublinear as visible in panel (c), showing that $S/T$ curves almost overlap over the $120-320\,{\rm K}$. Inset: a comparison between the dataset at $T=116.6\,{\rm K}$ is consistent with an electron mobility $\mu_e\approx11000\,{\rm cm^2/Vs}$ in the high carrier concentration limit.}
\end{center}
\end{figure}

\noindent which is valid for degenerate semiconductors in the low $T$ limit~\cite{note02,Mott}. Assuming that the temperature dependence of the mobility is small, this approximation would imply that $S/T$ is temperature-independent: data in Fig.~3c indicate that such expectation is almost exactly valid in the present case. We compare the model with the lowest temperature dataset ($T=116.6\,{\rm K}$) in the high band-filling regime (i.e. for large $\sigma$'s), which is the one we expect to best comply with the model approximations. Since the mobility is the only free parameter of the model (see Supplementary Materials), this leads to an interesting $\mu_e$ estimate, with respect to the more standard one $\mu_{e,F\!E}$ derived from field-effect in transport. In the inset to Fig. 3c we compare experimental data with predictions for an energy-independent $\mu_e=9000-13000\,{\rm cm^2/Vs}$, using no further adjustable parameters. A rather good agreement is obtained for $\mu_e\approx11000\,{\rm cm^2/Vs}$. Note that since the prediction is sensitive to the density of states, it is important to stress that only a minor non-parabolicity can be expected at the corresponding band filling (along the visible red part of the curve, $80$ to $120\,{\rm meV}$ from the $\Gamma$-point band edge). 

Albeit this is an approximate fit and lower conductivity data would require at least to take quantum confinement and finite temperature effects into account, the present $\mu_e$ estimate is robust and very instructive. In fact this value can be directly compared with $\mu_{e,F\!E}$, which we can determine in parallel starting from the $V_{bg}$ dependence of the NW transconductance (see Supplementary Materials). This procedure was observed to lead to values lower than $\approx 5000\,{\rm cm^2/Vs}$, for all explored $V_{bg}$ and $T$ regimes in the same NW. {Such a sizable underestimate of field-effect mobility with respect to our fit procedure is robust despite the relatively strong approximation of energy-independent $\mu_e$, since it is consistently observed across all explored filling regimes in the NWs, almost down to the pinch-off.} This discrepancy was never directly measured but was discussed independently in recent works, where it was pointed out that estimates of the NW free-carrier density values obtained by field effect are usually significantly overestimated as a consequence of surface-charge screening effects~\cite{Blomers12,Storm12} and non-ideal aspects of the gating geometry~\cite{Pitanti12}. In turn, this typically leads to a systematic underestimate of the carrier mobility. In this sense, the transport analysis discussed in Fig.~3 provides novel insight since it is largely independent from the influence of surface states. This is demonstrated by analyzing data in Fig.~4. Let us first examine $R(V_{bg})$ (panel (a)) and $V_{th}(V_{bg})$ (panel (b)) and compare data taken in the two opposite sweep directions at $\approx1\,{\rm V/min}$ with $T_D\approx325-326\,{\rm K}$ and $T_S\approx316-317\,{\rm K}$. No further parameter is modified during the measurement. A significant hysteresis is present in the $V_{bg}$ dependence: this behavior is a consequence of the slow dynamics of surface states and traps, leading to a time-dependent screening of the gate~\cite{Roddaro08,Karlstrom08}. Remarkably, despite the very large history-dependence of both $V_{th}$ and $R$ as a function of $V_{bg}$, the parametric trajectory in the $V_{th}$-$R$ space is largely unaffected by sweep direction as shown Fig. 4(c). This confirms nicely that even if hysteresis affects the free-charge filling of the NW as a function of $V_{bg}$, it has a negligible effect on the other transport parameters such as the carrier mobility $\mu_e$. 

\begin{figure}[h!]
\begin{center}
\includegraphics[width=0.48\textwidth]{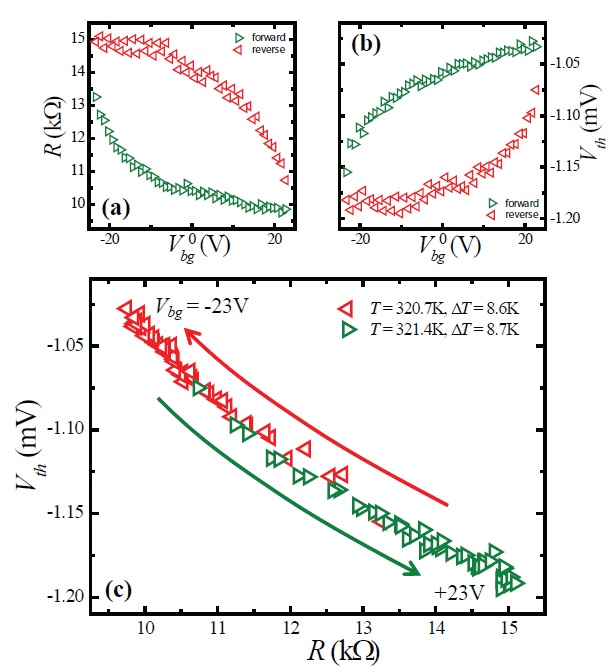}
\caption{$V_{th}(R)$, and thus the $S(\sigma)$, are largely insensitive from gating hysteresis caused by the filling/depletion of surface states and dielectric traps. The forward (green) and backward (red) sweeps $R(V_{bg})$ and $V_{th}(V_{bg})$ are reported in panels (a) and (b), and a strong hysteresis is highlighted. Differntly, the parametric dependence $V_{th}(R)$ in panel (c) is largely independent from the sweep direction. All measurements were performed at a relatively slow sweep rate $\approx 1\,{\rm V/min}$.}
\end{center}
\end{figure}

In conclusion, we demonstrated that large thermovoltages can be induced and tuned by gating in single-NW transistors. This result was obtained with a buried-heater scheme that makes it possible to impose both a large thermal bias $\Delta T$ and electric gating to the nanostructure. This experimental arrangement yielded a detailed mapping of $S(\sigma,T)$ that was compared with classic models for thermoelectric transport in degenerate semiconductors. The adopted approximations led to a alternative estimate of electron mobility $\mu_e\approx11000\,{\rm cm^2/Vs}$ in the NW. This value is significantly larger than the one obtain by field-effect, $\mu_{e,F\!E}$, on the same wire. The discrepancy can be understood in terms of the influence of slow surface- and trap-charge dynamics on field-effect mobility measurements, an influence that was shown to be negligible on $S$ vs. $\sigma$ dependence. The present results indicate that special care must be taken in the interpretation of transport results based on field-effects in these nanostructures.

{\bf Acknowledgements.} We gratefully acknowledge G. Signore for support with chemical passivation of the NWs. The work was partly supported by the Marie Curie Initial Training Action (ITN) Q-NET 264034 and by MIUR through the PRIN2009 project ``Dispositivi ad effetto di campo basati su nanofili e superconduttori ad alta temperatura critica''. SS acknowledges the support of the Academy of Finland, project NANoS (project
number 260815).

{\bf Methods.} {InAs NWs were grown by metal assisted chemical beam epitaxy in a Riber C-21 reactor, using tertiarybutylarsine (TBAs) and trimetylindium (TMIn) as metalorganic precursors for group V and group III elements with line pressures of 3.00 and 0.90 Torr, respectively. Ditertiarybutylselenide (DtBSe) with a line pressure of 0.30 Torr was used as n-doping source.} Devices were built starting from $80\,{\rm nm}$-diameter InAs NWs deposited by drop casting over a SiO$_2$/Si substrate. The substrate was heavily doped, with a resistivity $\rho=0.001-0.005\,{\rm \Omega\cdot cm}$, and was covered by a $280\,{\rm nm} $-thick oxide. In a first lithographic step aligned to the NW, heater contacts ($H_+$ and $H_-$ in Fig.~1) and the backgate voltage probes ($P_1$ and $P_2$) were defined by etching the SiO$_2$ layer and depositing a Ni/Au ($5\,{\rm nm}$/$25\,{\rm nm}$) bilayer on the exposed Si. In the second step, the thermometers and all connections to the bonding pads were defined by a single evaporation of a Ti/Au $10$/$100\,{\rm nm}$ bilayer.  Prior to evaporation the NW contact regions were exposed to an ${\rm NH_4S_x}$ solution to passivate the surface and avoid the formation of an insulating oxide layer~\cite{Suyatin}. The device was operated in a chamber containing a small amount of exchange gas (He), inside a variable-temperature cryostat. The sample holder was equipped with a calibrated Si-diode thermometer which was used to measure the bath temperature and calibrate the two resistive thermometers at the $S$ and $D$ contacts. Details about the heater operation and about the precise thermal and electrical calibration procedures adopted for the experiments are reported in the Supplementary Materials.


\begin{thebibliography}{99}

\bibitem{Majumdar04} A. Majumbdar, {\em Thermoelectricity in Semiconductor Nanostrucutres}, Science {\bf 303}, 777 (2004).

\bibitem{BookTE} G. S. Nolas, J. Sharp, H. J. Goldsmid, {\em Thermoelectrics: Basic Principles and New Materials Developments},  Springer New York, 2001.

\bibitem{Snyder08} G.~J. Snyder, and E.~S. Toberer, {\em Complex thermoelectric materials}, Nature Mat. {\bf 7}, 105 (2008).

\bibitem{Dresselhaus07} M.~S. Dresselhaus, G. Chen, M.~Y. Tang, R. Yang, H. Lee, D. Wang, Z. Ren, J.-P. Fleurial, and P. Gogna, {\em New Directions for Low-Dimensional Thermoelectric Materials}, Adv. Mat. {\bf 19}, 1043 (2007).

\bibitem{Vineis10} C.~J. Vineis, A. Shakouri, A. Majumdar, and M.~G. Kantzidis, {\em Nanostructured Thermoelectrics: Big Efficiency Gains from Small Features}, Adv. Mat. {\bf 22}, 3970 (2010).

\bibitem{Shi12} L. Shi, {\em Thermal and Thermoelectric Transport in Nanostructures and Low-Dimensional Systems}, Nanoscale and Microscale Thermophysical Engineering {\bf 16}, 79 (2012).

\bibitem{Venkatasubramanian01} R. Venkatasubramanian, E. Siivola, T. Colpitts and B. O'Quinn, {\em Thin-film thermoelectric devices with high room-temperature figures of merit}, Nature {\bf 413}, 597 (2001).

\bibitem{Boukai08} A.~I. Boukai, Y. Bunimovich, J. Tahir-Kheli, J.-K. Yu, W.~A. Goddard III, and J.~R. Heath, {\em Silicon nanowires as effcient thermoelectric materials}, Nature {\bf 451}, 06458 (2008).

\bibitem{Poudel08} B. Poudel, Q. Hao, Y. Ma, Y. Lan, A. Minnich, B. Yu, X. Yan, D. Wang, A. Muto, D. Vashee, X. Chen, J. Liu, M.~S. Dresselhaus, G. Chen, and Z. Ren, {\em High-Thermeolectric Performance of Nanostructured Bismuth Antimony Telluride Bulk Alloys}, Science {\bf 320}, 634 (2008)

\bibitem{Heremans08} J.~P. Heremans, V. Jovovic, E.~S. Toberer, A. Saramat, K. Kurosaki, A. Charoenphkdee, S. Yamanaka, and G.~S. Snyder, {\em Enhancement of Thermoelectric Efficiency in PbTe by Distortion of the Electronic Density of States}, Science {\bf 321}, 554 (2008).

\bibitem{Shi10} L. Shi, D. Yao, G. Zhang, and B. Li, {\em Large thermoelectric figure of merit in Si$_{1-x}$Ge$_x$ nanowires}, Appl. Phys. Lett. {\bf 96}, 173108 (2010).

\bibitem{Martinez11} J.~A. Martinez, P.~P. Provencio, S.~T. Picraux, J.~P. Sullivan, and B.~S. Swartzentruber, {\em Enhanced thermoelectric figure of meriti in SiGe alloy nanowires by boundary and hole-phonon scattering}, J. Appl. Phys. {\bf 110}, 074317 (2011).

\bibitem{Lee11}, S.~H. Lee, W. Shim, S.~Y. Jang, J.~W. Roh, P. Kim, J. Park, and W. Lee, {\em Thermeoelctric properties of inidividual single-crystalline PbTe nanowires grown by a vapour transport method}, Nanotech. {\bf 22}, 295707 (2011).

\bibitem{Zhou11} F. Zhou, A.~L. Moore, J. Bolinsson, A. Persson, L. Fr\"oberg, M.~T. Pettes, H. Kong, L. Rabenberg, P. Caroff, D.~A. Stewart, N. Mingo, K.~A. Dick, L. Samuelson, H. Linke, and L. Shi, {\em Thermal conductivity of indium arsenide nanowires with wurtzite and zinc blende phases}, Phys. Rev. B {\bf 83}, 205416 (2011).

{
\bibitem{Tian12} Y. Tian, M.~R. Sakr, J.~M. Kinder, D. Liang, M.~J. MacDonald, R.~L.~J. Qiu, H.-J. Gao, and X.~P.~A. Gao, {\em One-Dimensional Quantum Confinement Effect Modulated Thermoelectric Properties in InAs Nanowires}, Nano Lett. {\bf 12}, 6492 (2012).

\bibitem{Moon13} J. Moon, J.-H. Kim, Z.~C.~Y. Chen, J. Xiang, R. Chen, {\em Gate-Modulated Thermoelectric Power Factor of Hole Gas in Ge-Si Core-Shell Nanowires}, Nano Lett. {\bf 13}, 1196 (2013).
}

\bibitem{RoddaroNR11} S. Roddaro, A. Pescaglini, D. Ercolani, L. Sorba, F. Giazotto, and F. Beltram, {\em Hot-electron Effects in InAs Nanowire Josephson Junctions}, Nano Res. {\bf 4}, 259 (2011).

\bibitem{Spathis11} P. Spathis, S. Biswas, S. Roddaro, L. Sorba, F. Giazotto, and F. Beltram, {\em Hybrid InAs nanowire-vanadium proximity SQUID}, Nanotech. {\bf 22}, 105201 (2011).

\bibitem{Giazotto11} F. Giazotto, P. Spathis, S. Roddaro, S. Biswas, F. Taddei, M. Governale, and L. Sorba, {\em A Josephson quantum electron pump}, Nature Phys. {\bf 7}, 857 (2011).

\bibitem{Vitiello12} M.~S. Vitiello, D. Coquillat, L. Viti, D. Ercolani, F. Teppe, A. Pitanti, F. Beltram, L. Sorba, W. Knap, and A. Tredicucci, {\em Room-Temperature Terahertz Detectors Based on Semiconductor Nanowire Field-Effect Transistors} Nano Lett. {\bf 12}, 96 (2012).

\bibitem{Pitanti11} A. Pitanti, D. Ercolani, L. Sorba, S. Roddaro, F. Beltram, L. Nasi, G. Salviati, and A. Tredicucci, {\em InAs/InP/InSb Nanowires as Low Capacitance n-n Heterojunction Diodes} Phys. Rev. X {\bf 1}, 011006 (2011).

\bibitem{ThelanderAPL2003} C. Thelander, T. M\aa{}rtensson, M.~T. Bj\"ork, B.~J. Ohlsson, M.~W. Larsson, L.~R. Wallenberg, and L. Samuelson, {\em Single-electron transistors in heterostructure nanowires}, Appl.Phys.Lett. {\bf 83}, 2052 (2003).

\bibitem{Bjork04} M.~T. Bj\"ork, C. Thelander, A.~E. Hansem, L.~E. Jensen, M.~W. Larsson, L.~R. Wallenberg, and L. Samuelson, {\em Few-Electron Quantum Dots in Nanowires}, Nano Lett. {\bf 4}, 1621 (2004).

\bibitem{Fuhrer07} A. Fuhrer, L.~E. Fr\"oberg, J.~N. Pedersen, M.~W. Wallenberg, A. Wacker, M.-E. Pistol, and L. Samuelson, {\em Few-Electron Double Dots in InAs/InP Nnaowire Heterostrucutres}, Nano Lett. {\bf 7}, 243 (2007).

\bibitem{Roddaro11}  S. Roddaro, A. Pescaglini, D. Ercolani, L. Sorba, and F. Beltram, {\em Manipulation of Electron Orbitals in Hard-Wall InAs/InP Nanowire Quantum Dots}, Nano Lett. {\bf{11}}, 1695 (2011).

\bibitem{Romeo12}  L. Romeo, S. Roddaro, A. Pitanti, D. Ercolani, L. Sorba, and F. Beltram, {\em Electrostatic Spin Control in InAs/InP Nanowire Quantum Dots}, Nano Lett. {\bf 12}, 4490 (2012).

\bibitem{Hoffmann09} E.~A. Hoffmann, H.~A. Nilsson, J.~E. Matthews, N. Nakpathomkun, A.~I. Persson, L. Samuelson, and H. Linke, {\em Measuring Temperature Gradients over Nanometer Length Scales}, Nano Lett. {\bf 9}, 779 (2009).

\bibitem{Zhang11} Y. Zhang, M.~S. Dresselhaus, Y. Shi, Z. Ren, and G. Chen, {\em High Thermoelectric Figure-of-Merit in Kondo Insulator Nanowires at Low Temperatures} Nano Lett. {\bf 11}, 1166 (2011).

\bibitem{Halpern12} E. Halpern, G.Elias, A.~V. Kretinin, H. Shtrikman, and Y. Rosenwaks, {\em Direct measurement of surface states density and energy distribution in individual InAs nanowires}, Appl. Phys. Lett. {\bf 100}, 262105 (2012).

%

%
\bibitem{Mott} N.~F. Mott and H. Jones, {\em The Theory of the Properties of Metals and Alloys}, Courier Dover Publications, 1985.

%
\bibitem{note02} The expected energy spacing between the one-dimensional subbands can be calculated to be $5-10\,{\rm meV}$ for our wires, the adoption of a three-dimensional band dispersion relation provides an acceptable description of the nanowire at such high carrier densities.

\bibitem{Blomers12} Ch. Bl\"omers, T.Grap, M.~I. Lepsa, J. Moers, St. Trellenkamp, D. Gr\"utzmacher, H. L\"uth, and Th. Sch\"apers, {\em Hall effect measurement on InAs nanowires}, Appl. Phys. Lett. {\bf 101}, 152106 (2012).

\bibitem{Storm12} K. Storm, F. Halvardsson, M. Heurlin, D. Lingren, A. Gustafsson, P.~M. Wu, B. Monemar, and L. Samuelson, {\em Spatially resolved Hall effect measurement in a single semiconductor nanowire}, Nature Nanotech. {\bf 7}, 718 (2012).

\bibitem{Pitanti12} A. Pitanti, S. Roddaro, M.S.Vitiello, and A.Tredicucci, {\em Contacts shiedling in nanowire field effect transistors} J. Appl. Phys. {\bf 111}, 064301 (2012).

\bibitem{Roddaro08} S. Roddaro, K. Nilsson, G. Astromskas, L. Samuelson, L.-E. Wernersson, O. Karlstr\"om, and A. Wacker, {\em InAs nanowire metal-oxide-semiconductor capacitors}, Appl. Phys. Lett. {\bf 92}, 253509 (2008). 

\bibitem{Karlstrom08} O. Karlstr\"om, A. Wacker, K. Nilsson, G. Astromskas, S. Roddaro, L. Samuelson and L.-E. Wernersson, , {\em Analysing the capacitance-voltage measurements of vertical wrapped-gated nanowires}, Nanotech. {\bf 19}, 435201 (2008). 

\bibitem{Suyatin} D.~B. Suyatin, C. Thelander, M.~T. Bj\"{o}rk, I. Maximov, and L. Samuelson, {\em Sulfur passivation for ohmic contact formation to InAs nanowires}, Phys. Rev. B {\bf 44}, 1646 (1991).




\end{thebibliography}
\end{document}